\documentclass[10pt,aps,prb,twocolumn,showpacs,superscriptaddress,floatfix]{revtex4-2}

\usepackage{graphicx}
\usepackage{dcolumn}
\usepackage{bm}
\usepackage{amsmath}
\usepackage{amssymb}
\usepackage{tabularx}
\usepackage{graphicx}
\usepackage{dcolumn}
\usepackage{bm}
\usepackage{amssymb}
\usepackage{siunitx}
\usepackage{textalpha}
\usepackage{xcolor}
\usepackage{soul}
\usepackage{epstopdf}
\usepackage[export]{adjustbox}
\usepackage{hyperref}
\DeclareUnicodeCharacter{2212}{\ensuremath{-}}

\begin{document}

\title{Multiple Field-Induced Transitions, Magnetocaloric Response and Anisotropic Magnetotransport in Orthorhombic RAlGe (R = Dy, Ho)}

\author{Ram Kumar}
\email{ramphy21@umd.edu}
\affiliation {Maryland Quantum Materials Center, Department of Physics, University of Maryland, College Park, MD 20742, USA}
\author{D. S. Sokratov}
\affiliation {Maryland Quantum Materials Center, Department of Physics, University of Maryland, College Park, MD 20742, USA}
\author{P. Sobel}
\affiliation {Maryland Quantum Materials Center, Department of Physics, University of Maryland, College Park, MD 20742, USA}
\author{C. Mueller}
\affiliation {Maryland Quantum Materials Center, Department of Physics, University of Maryland, College Park, MD 20742, USA}
\author{D. E. Graf}
\affiliation {National High Magnetic Field Laboratory, Tallahassee, Florida 32310, USA}
\author{J. Paglione}
\email{paglione@umd.edu}
\affiliation {Maryland Quantum Materials Center, Department of Physics, University of Maryland, College Park, MD 20742, USA}
\affiliation {Canadian Institute for Advanced Research, Toronto, ON M5G-1Z8, Canada}

\date{\today}

\begin{abstract}
We present a comprehensive study of the magnetic, thermodynamic, and magnetocaloric properties of single-crystalline DyAlGe and HoAlGe, both members of the RAlGe Weyl semimetal family. Magnetization, heat capacity and transport measurements reveal multiple magnetic phase transitions associated with the evolution of complex magnetic structures. For DyAlGe, the primary magnetic ordering occurs at \textit{$T_{N1}$}= 30.5 K , while HoAlGe orders at a much lower temperature of \textit{$T_{N1}$}= 11.5 K, followed by additional first-order magnetic transitions at 6.6 K and 8 K, respectively. The magnetization isotherms display pronounced anisotropy for the B $\parallel$ a, consistent with competing exchange and crystal-field interactions. The magnetic entropy change ($\Delta S_M$) derived from isothermal magnetization demonstrates a significant magnetocaloric effect (MCE) with sign change in both systems, with a maximum $\Delta S_M$ of $\approx$ 5$-$23 J/kg-K along B $\parallel$ a for a field change of 0$-$7 T. The sign reversal of $\Delta S_M$ across field-induced transitions highlights the complex field-tuned spin configurations. The moderate MCE and magnetic anisotropy suggest that these materials may serve as model systems for studying magnetocaloric effects in topologically nontrivial intermetallics with interplay between localized 4\textit{f} magnetism and crystal symmetry.

\end{abstract}

\maketitle

\section{Introduction}
Weyl semimetals exhibit a variety of remarkable physical properties, including topologically protected electronic states that coexist with conventional metallic conduction, giving rise to unconventional transport phenomena. Recently, magnetic topological materials in the family of strongly correlated \textit{f}-electrons intermetallics have attracted much attention as they provide rich physical properties due to the interplay between correlated-electron phenomena and topological electronic bands \cite{Bernevig2022}, which make them valuable for both fundamental research and technological applications \cite{KHJBuschow_1977, RevModPhys.81.1551}. More recently, the rare-earth based RAlGe (R = rare earth) intermetallics have attracted considerable attention as potential Weyl semimetal candidates \cite{PhysRevB.84.235126,PhysRevB.97.041104, doi:10.1126/science.1245085, Neupane2014, doi:10.1021/ic00025a021}. The RAlGe family provides a platform for investigating the interplay between rare-earth magnetism and nontrivial electronic structures. While the electronic and magnetic properties of several members have been explored, the magnetic phase diagrams and anisotropic transport responses of the heavy rare-earth compounds remain comparatively less understood. In particular, a systematic investigation of single-crystalline DyAlGe and HoAlGe is important for establishing how rare-earth-dependent magnetic interactions and crystal-electric-field effects influence their field-induced transitions and magnetocaloric properties.\\

Most compounds in the RAlGe family exhibit intriguing magnetic and transport properties arising from their diverse electronic structures and complex magnetic ordering. The light rare-earth-based RAlGe (R = La-Eu) crystallize in a disordered centrosymmetric body-centered tetragonal structure \cite{doi:10.1021/ic00025a021, PUKAS2004162}; however, their structural model is corrected as the ordered noncentrosymmetric I4$_1$md type structure in recent studies \cite{Klepp:a21087}. The Ce- and Pr-based members of the RAlGe family exhibit intriguing magnetic and electrical properties \cite{PhysRevB.98.245132, PUKAS2004162, FLANDORFER1998191,doi:10.1126/science.aat0348,Singh02072020, 10.1063/1.5090795, Destraz2020, Sanchez2020, 10.1063/1.5132958}. Single-crystalline NdAlGe, in particular, displays a field-induced metamagnetic transition, a large anomalous Hall effect, and a helical magnetic structure\cite{Zhao_2022, PhysRevB.107.224414, PhysRevMaterials.7.034202}. Recent studies show that the heavy rare-earth-based compounds RAlGe (R = Y-Lu) crystallize in the YAlGe-type structure with space group $Cmcm$ \cite{WANG2021167739, WANG2022163623}, where the Al atoms occupy the 4a(0, 0, 0) site and R and Ge atoms occupy two kinds of inequivalent 4c (0, y, 0.25) sites, respectively. Relatively little attention has been devoted to the heavy rare-earth members of the RAlGe family. Until recently, most of these compounds had been investigated primarily in polycrystalline form \cite{doi:10.1021/jpcc.1c07073,WANG2021167739,WANG2022163623}. More recently, studies of single-crystalline GdAlGe and TbAlGe have begun to shed light on their structural, magnetic, and electronic properties. Orthorhombic GdAlGe has been reported to exhibit Dirac-like linear band dispersion \cite{r7gz-kwqw}, while high-field measurements on TbAlGe have revealed complex magnetic-field-induced magnetotransport behavior\cite{z39k-8vrq}. These studies, together with recent investigations of the heavy rare-earth RAlGe series\cite{SLADE2026174537}, provide a basis for exploring the magnetic anisotropy, field-induced transitions, magnetocaloric and magnetotransport properties of DyAlGe and HoAlGe.

\begin{figure}[h]
	\adjincludegraphics[height=11.8cm,trim={1.2cm 0.3cm 0.5cm 0cm},clip]{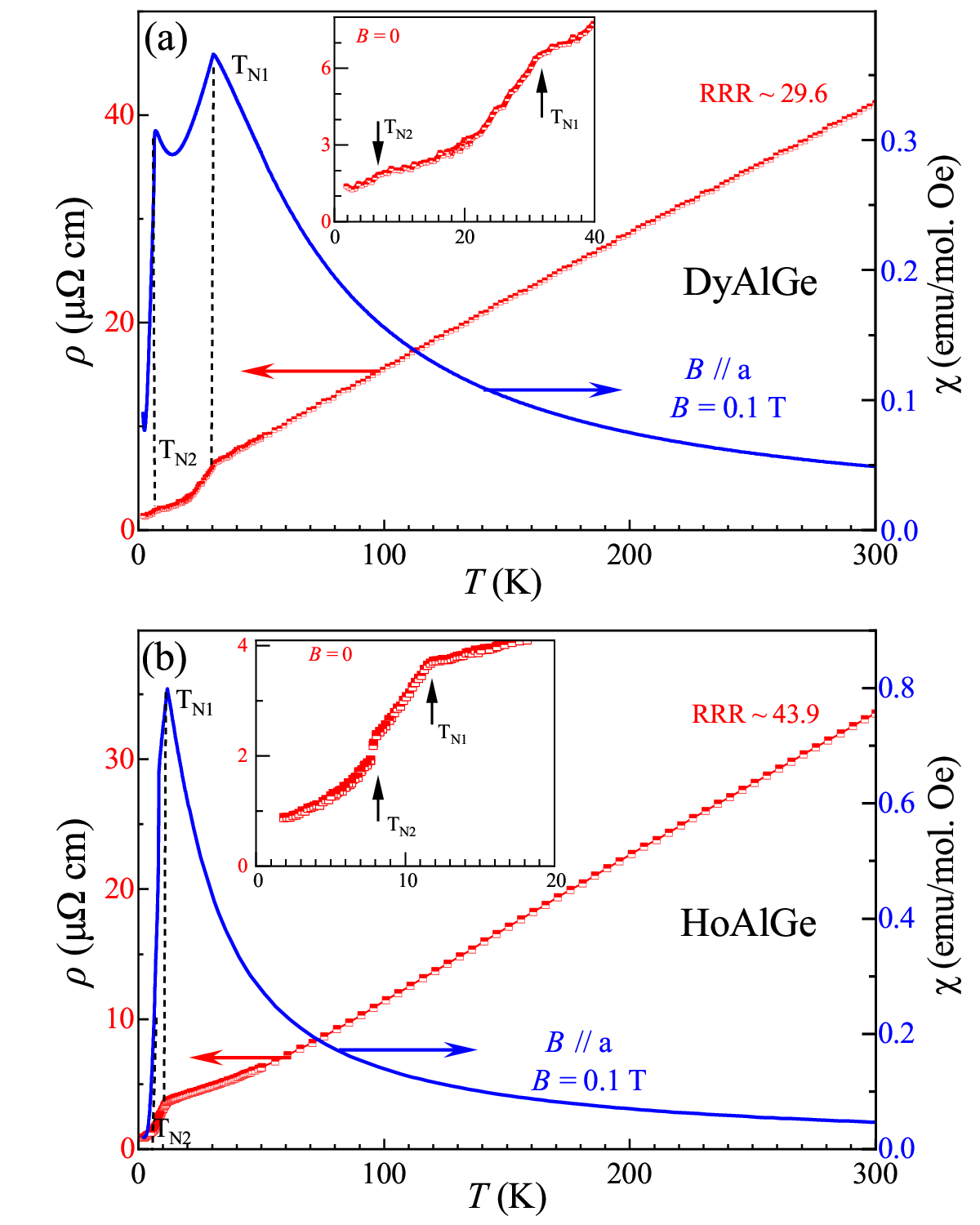}
	\caption{Temperature dependence of the zero-field electrical resistivity, $\rho$(T), and magnetic susceptibility, $\chi(T)$, for $B \parallel a$ in (a) DyAlGe and (b) HoAlGe. The residual resistivity ratios, $\mathrm{RRR} = \rho(300\,\mathrm{K})/\rho_0$, are indicative of high-quality single crystals.}
    \label{fgr:RRR}
\end{figure}

In this study, we present the comprehensive investigation of the magnetic, magnetocaloric, and high-field magnetotransport properties of single-crystalline orthorhombic DyAlGe and HoAlGe, including measurements in magnetic fields up to 35 T. Through detailed characterization of these single crystals, we investigate the evolution of the two antiferromagnetic (AFM) states, which emerge below $T_{N1}$ = 30.5~K and $T_{N2}$ = 6.6~K in DyAlGe and $T_{N1}$ = 11.5~K and $T_{N2}$ = 8~K in HoAlGe (Fig.~\ref{fgr:RRR}), and track their evolution under applied magnetic fields, revealing a rich landscape of field-induced metamagnetic transitions. We further elucidate the pronounced magnetic anisotropy of these orthorhombic compounds and discuss the possible roles of anisotropic exchange interactions, crystal-electric-field effects, and spin–orbit coupling in shaping their complex magnetic structures. In addition, the substantial magnetocaloric response highlights the potential of these materials for magnetic-refrigeration applications. Together, these findings deepen our understanding of the interplay between localized rare-earth magnetism and electronic transport in the RAlGe family, a class of materials associated with Weyl-semimetal physics.

\section{EXPERIMENTAL DETAILS}

Single crystals of DyAlGe and HoAlGe were grown using the high-temperature self-flux technique with molten Al as a solvent. Dy and Ho with 99.9$\%$ purity, Al (shot) and Ge (lump) with purity greater than 99.999$\%$, were obtained from Alfa-Aesar. The reactions were carried out in 2-cm$^3$ alumina crucibles, which were encapsulated in evacuated fused silica jackets by flame sealing. The temperature profile: ramping to 1100~$^\circ$C with the rate 100$^{\circ}$C$/$hr, homogenization for up to 24~hr, cooling to 850~$^\circ$C at 2$^{\circ}$C/hr rate. At 850~$^\circ$C, the molten Al was removed by centrifugation. The as-grown crystal is shown in the inset of Fig.~\ref{fgr:XRD}(b).  The phase purity of these crystals was checked with an X-ray diffractometer (Rigaku, MiniFlex600), where X-ray diffraction (XRD) patterns were recorded with monochromated Cu-\textit{K}\(\alpha\) radiation ($\lambda \sim 1.5406~\AA$), and crystal orientation along principal crystallographic directions was determined by Laue diffraction using a polychromatic X-ray source. The Rietveld refinement technique \cite{Rietveld:a07067} using FullProf software packages has been used for phase identification. Further, the chemical composition is verified using a scanning electron microscope (SEM) with an energy-dispersive X-ray spectroscopy (EDS). The magnetic measurements were carried out using a SQUID-VSM, Quantum Design, and Dynacool with VSM-mode for magnetic field up to 14~T.

\section{Results and Discussion}
\subsection{Structural properties}
\begin{figure}[h]
	\adjincludegraphics[height=12.5cm,trim={0cm 0cm 0cm 0cm},clip]{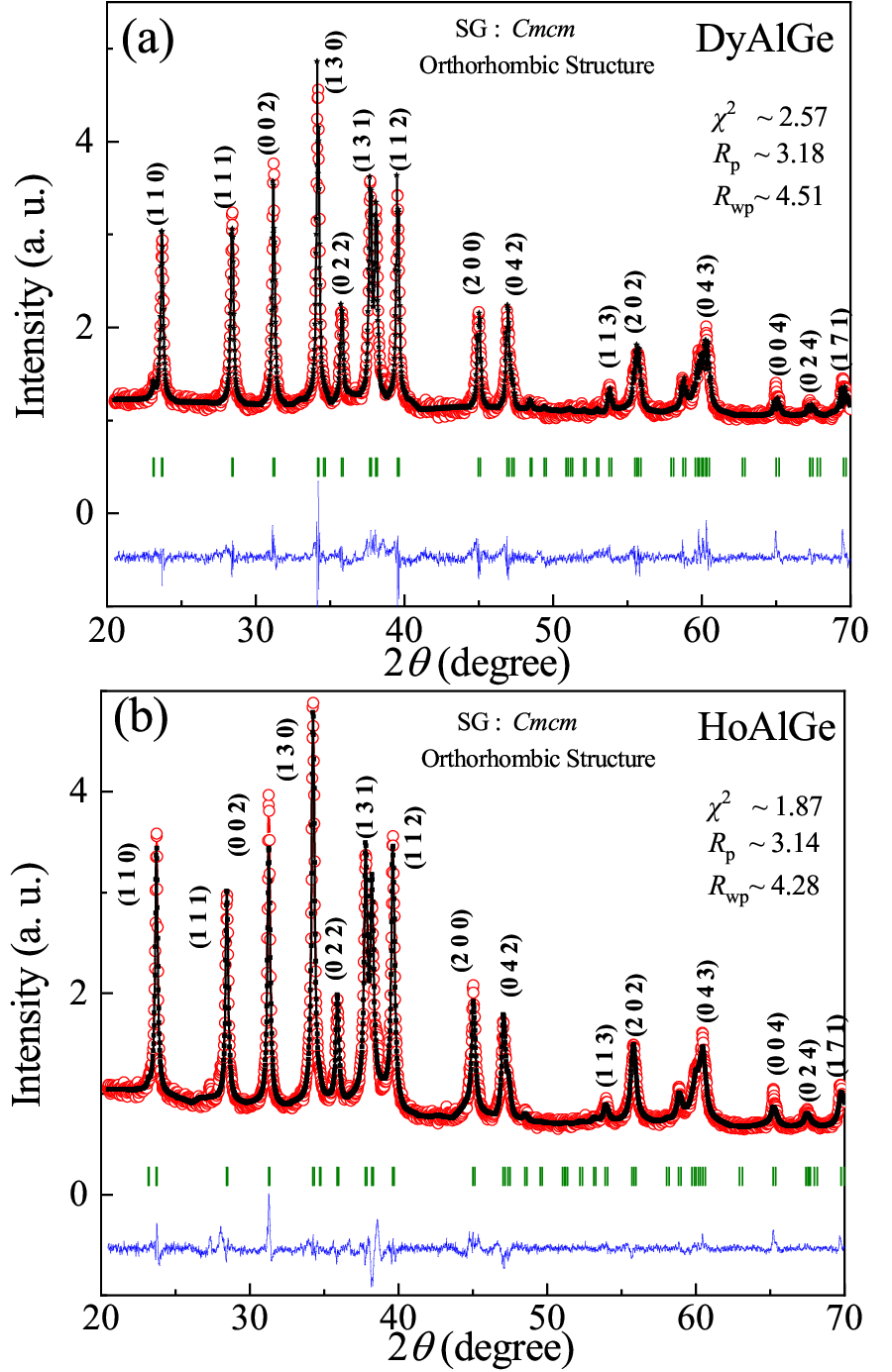}
	\caption{Structural analysis of both DyAlGe and HoAlGe crystals. Panels (a) and (b) represent Rietveld refinement of powder X-ray diffraction data with the experimental data (Red symbols) and the black line representing the calculated data. A set of vertical green bars represents the Bragg peak positions of the orthorhombic (Space group: $Cmcm$) structure, and the blue line at the bottom represents the difference between the observed and calculated diffraction patterns, for DyAlGe and HoAlGe, respectively.}
	\label{fgr:XRD}
\end{figure}

Fig.~\ref{fgr:XRD}(a) shows the Rietveld refinement of the powder XRD data obtained at room temperature of ground single crystals of DyAlGe and HoAlGe. Structural analysis confirms that those DyAlGe and HoAlGe crystals are in the orthorhombic structural phase, similar to the recently reported TbAlGe\cite{z39k-8vrq}. Detailed powder XRD refinements converge in the space group $Cmcm$ with $a$- $b$- and $c$-axis lattice parameters 4.028(1)~Å, 10.369(1)~Å and 5.734(1)~Å, respectively, for DyAlGe, and 4.025(1)~Å, 10.335(1)~Å and 5.719(1)~Å, respectively, for HoAlGe which correspond well to previous polycrystalline studies \cite{WANG2022163623}. The decrease in the lattice parameters from DyAlGe to HoAlGe is consistent with the expected lanthanide contraction across the rare-earth series. EDS performed on the cleaned surfaces of the single crystal yielded an atomic composition of 31.97(1)\% Dy, 32.47(2)\% Al, and 35.56(1)\% Ge for DyAlGe, and 31.21(1)\% Ho, 33.85(2)\% Al, and 34.94(1)\% Ge for HoAlGe. These values are in good agreement with the compositions obtained from XRD refinement.


\subsection{Magnetic properties}
\subsubsection{\textbf{DyAlGe}}
\begin{figure}[t]
	\adjincludegraphics[height=16cm,trim={0.6cm 0.25cm 0cm 0 cm},clip]{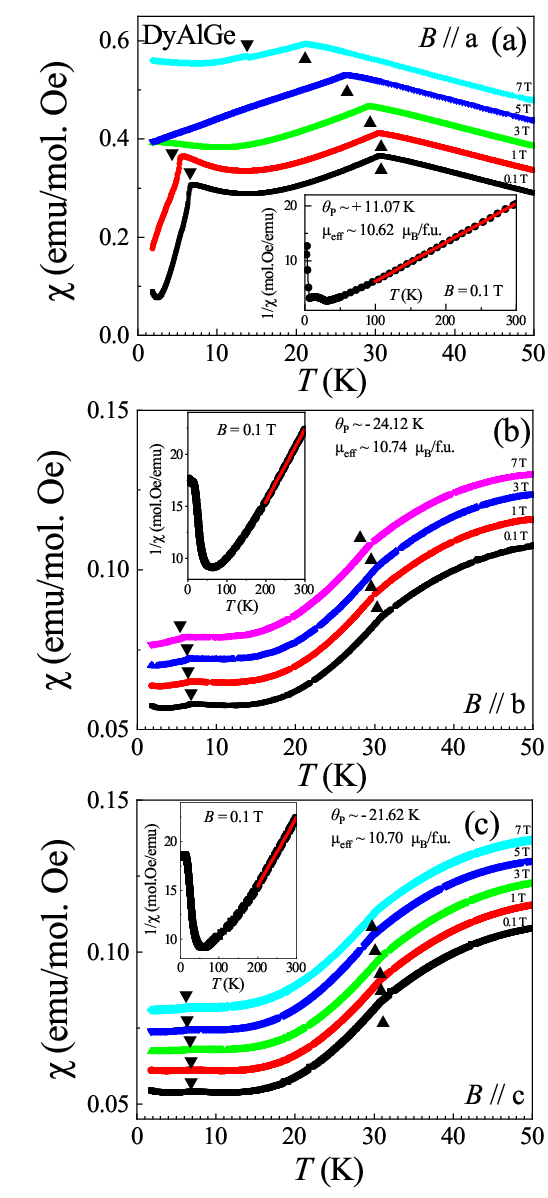}
\caption{Temperature dependence of magnetic susceptibility $\chi{_{dc}}$(\textit{T}) of DyAlGe measured under a magnetic field from 0.1~T to 7~T in all three $B \parallel a$, $B \parallel b$, and $B \parallel c$ directions as shown in fig.~2~(a), (b), and (c), respectively. The upward and downward triangles indicate \textit{$T_{N1}$} and \textit{$T_{N2}$}, respectively. The insets show the inverse susceptibility plots. For clarity, the curves measured at fields above 0.1 T are vertically offset to better illustrate the magnetic transitions under applied magnetic fields.}
	\label{fgr:MvsT_DyAlGe}
\end{figure}

\begin{figure}[h]
	\adjincludegraphics[height=16cm,trim={0.3cm 0cm 0.5cm 0cm},clip]{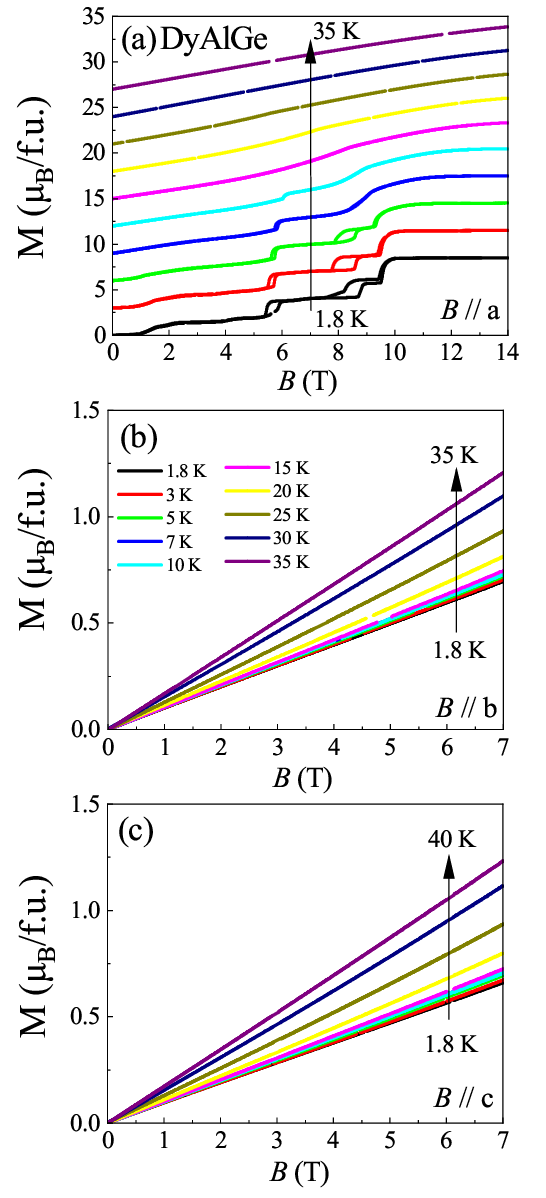} 
    \caption{Magnetic field (\textit{B})dependence of magnetization (\textit{M}) of DyAlGe measured at different temperatures for the applied magnetic field parallel to three principal crystallographic directions, i.e., (a) \textit{$B \parallel a$}, (b) \textit{$B \parallel b$}, and (c) \textit{$B \parallel c$}, respectively. We note that for \textit{$B \parallel a$}, there are multiple metamagnetic transitions, while in contrast, the other field directions are mostly featureless with minor curvature. The curves in fig.~3~(a) are shifted upward for clarity to illustrate the multiple metamagnetic transitions along with hysteresis.}
	\label{fgr:MvsB_DyAlGe}
\end{figure}

Fig.~\ref{fgr:MvsT_DyAlGe} presents the temperature-dependent \textit{dc}-magnetic susceptibility $\chi$(\textit{T}) of DyAlGe measured in the ZFC mode with different applied fields along the three principal crystallographic axes of the orthorhombic structure of the system. It is observed that $\chi$(\textit{T}) exhibits two phase transitions of AFM at Néel temperatures of \textit{$T_{N1}$}= 30.5~K and \textit{$T_{N2}$}= 6.6~K. The Curie-Weiss fits of inverse susceptibilities in the paramagnetic (PM) region for magnetic fields parallel to all three principle axis is shown as the respective insets of Fig.~\ref{fgr:MvsT_DyAlGe}. The effective magnetic moment determined experimentally is comparable to the theoretical free-ion value of 10.63 $\mu$$_{B}$ for the Dy$^{3+}$ ion. Similar to the other compounds of this series, the $a$-axis, which is the orthorhombic axis of these systems\cite{z39k-8vrq}, looks to have a more interesting response. The negative values of $\theta_{\mathrm{P}}$ obtained from the Curie-Weiss fits for $B \parallel b$ and $B \parallel c$ indicate that antiferromagnetic exchange interactions dominate along these directions, whereas the positive $\theta_{\mathrm{P}}$ for $B \parallel a$ suggests predominant ferromagnetic exchange. This anisotropic behavior highlights the strong magnetocrystalline (single-ion) anisotropy of the Dy ions \cite{PhysRevB.100.094442, PhysRevB.99.100406}, arising from the directional dependence of the exchange interactions in the system. Furthermore, the magnetic susceptibility exhibits pronounced anisotropy with respect to the applied field direction. A broad hump centered around 50 K is observed for $B \parallel b$ and $B \parallel c$, whereas no comparable feature is present for $B \parallel a$. The appearance of this broad maximum well above $T_{\mathrm{N}}$ is indicative of short-range magnetic correlations, likely enhanced by crystal electric field effects associated with the Dy$^{3+}$ ions. Next, to gain a deeper insight into the behavior of these anisotropic magnetic phase transitions, we have measured $\chi$(\textit{T}) with different applied magnetic fields up to 7~T along all three directions. The $\chi$(\textit{T}) plots with applied magnetic fields for $B \parallel a$ look distinctive, as shown in Fig.~\ref{fgr:MvsT_DyAlGe}(a). There is a drastic change in the magnetic transition, particularly for $T_{N1}$, which shifted from 30~K (for $B$ = 0.1~T) to $\sim$20~K ( for $B$ = 7~T). There is not much variation in $T_{N1}$ and $T_{N2}$ for the cases $B \parallel b$ and $B \parallel c$ [Fig.~\ref{fgr:MvsT_DyAlGe}(b) and (c)]. Note that the Néel temperature of DyAlGe ($T_{N1}$ = 30~K) is markedly higher than the value expected from de Gennes scaling based on the Gd analog ($T_{N}$ = 30~K) \cite{r7gz-kwqw}. This enhancement indicates a breakdown of simple RKKY scaling, likely caused by strong anisotropic exchange interactions and crystal-field effects \cite{MAJUMDAR2001645}. Such behavior underscores the complex interplay between localized 4$f$ magnetism and itinerant electron topology in DyAlGe.                                                    
To further investigate the effect of the magnetic field, we have measured isothermal magnetization at selected temperatures up to 35 K. Similar to the $\chi(T)$ response, multiple metamagnetic transitions are observed for B $\parallel$ a. Fig.~\ref{fgr:MvsB_DyAlGe}(a) shows magnetization as a function of the field at different temperatures between 1.8-35~K for B $\parallel$ a. Multiple metamagnetic transitions (MMTs) at the lowest measured temperature (1.8~K) with significant hysteresis are observed on the slopes of the $M(B)$ curves near 1~T and 6~T. This first-order-like magnetic transition gradually gets smeared with increasing temperature and shows reversible up and down $M(B)$ curves above 7 K. The transition observed around 5.5~T shifts to higher fields with increasing temperature, suggesting the presence of more intriguing magnetic behavior at further higher field values. The $M(B)$ curves for other directions are observed to be predominantly reversible with magnetic field directions without any field-induced transition. Further, $M(B)$ curves do saturate around 10~T field at 1.8~K, and the spontaneous magnetic moment at 1.8~K is ~9$\mu$$_{B}$/Dy$^{3+}$ (~0.5$\mu$$_{B}$/Dy$^{3+}$r up to 7 T for other directions), which is much lower than the free Dy$^{3+}$ magnetic moment which hints a modulated/chiral magnetic structure due to competing anisotropic exchange interaction with a non-trivial magnetic ground state. 

\begin{figure}[ht]
	\adjincludegraphics[height=16cm,trim={0cm 0.25cm 0cm 0 cm},clip]{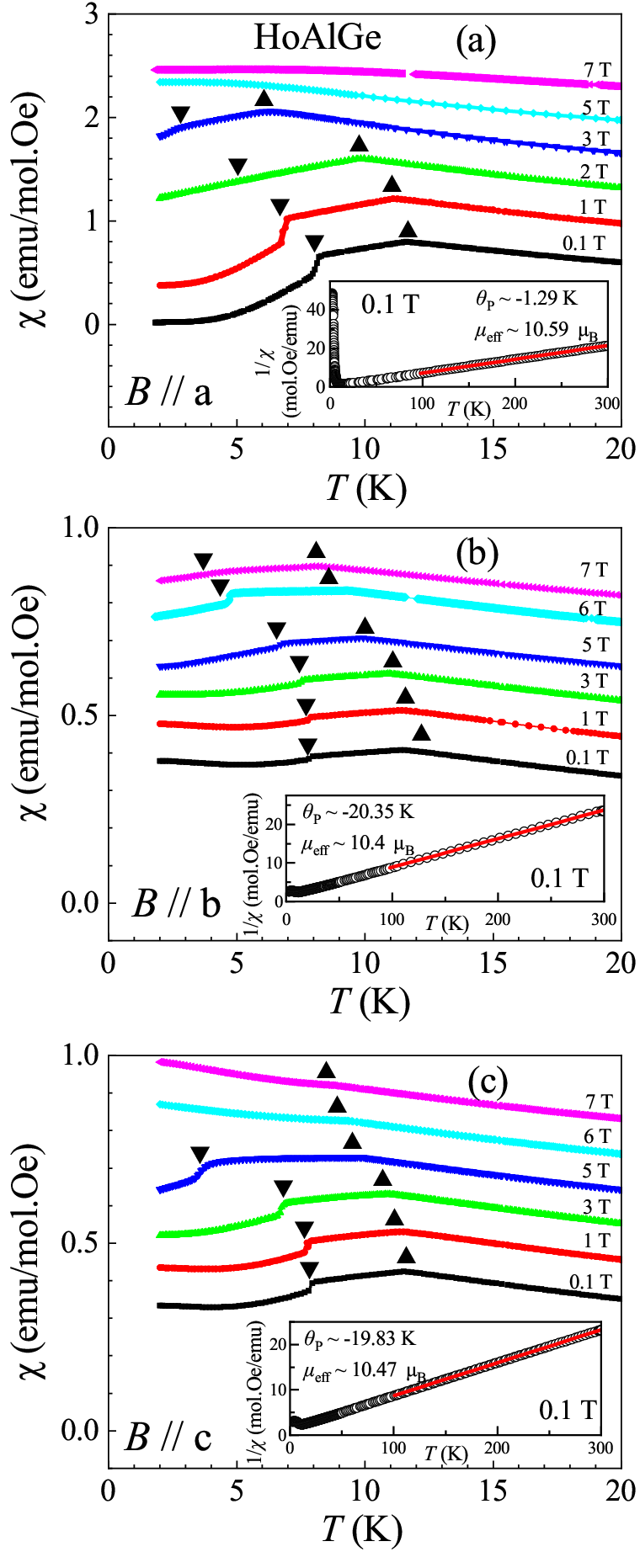}
\caption{Temperature dependence of magnetic susceptibility $\chi{_{dc}}$(\textit{T}) of HoAlGe measured under a magnetic field from 0.1~T to 7~T in all three $B \parallel a$, $B \parallel b$, and $B \parallel c$ directions as shown in fig.~2~(a), (b), and (c), respectively. The upward and downward triangles indicate $T_{N1}$ and $T_{N2}$, respectively. The insets show the inverse susceptibility plots. For clarity, the curves measured at fields above 0.1 T are vertically offset to better illustrate the magnetic transitions under applied magnetic fields.}
	\label{fgr:MvsT_Ho}
\end{figure}


\begin{figure}[h]
	\adjincludegraphics[height=16cm,trim={0cm 0cm 0cm 0cm},clip]{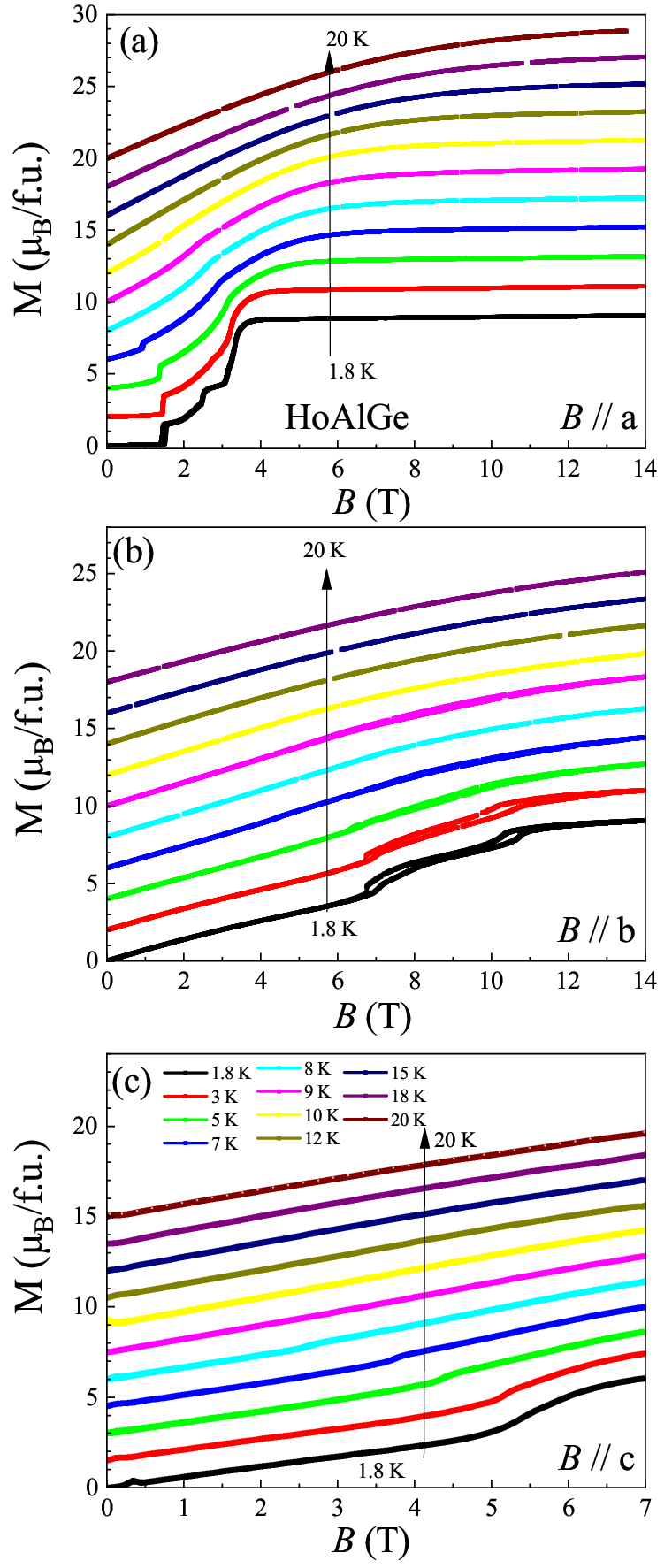}
    \caption{Magnetic field (\textit{B}) dependence of magnetization (\textit{M}) of HoAlGe measured at different temperatures for the applied magnetic field parallel to three principal crystallographic directions, i.e., (a) \textit{$B \parallel a$}, (b) \textit{$B \parallel b$}, and (c) \textit{$B \parallel c$}, respectively. We note that for \textit{$B \parallel a$}, there are multiple metamagnetic transitions, in contrast to DyAlGe, the other field orientations in HoAlGe also exhibit metamagnetic transitions. All the curves above 1.8 K, are shifted upward for clarity to illustrate the multiple metamagnetic transitions along with hysteresis.}
	\label{fgr:MvsB_Ho}
\end{figure}


\subsubsection{\textbf{HoAlGe}}
Fig.~\ref{fgr:MvsT_Ho} presents the temperature-dependent \textit{dc}-magnetic susceptibility $\chi$(\textit{T}) of HoAlGe measured in the ZFC mode with different applied fields along the three principal crystallographic axes of the orthorhombic structure of HoAlGe. It is observed that similar to DyAlGe, $\chi$(\textit{T}) exhibits two phase transitions of AFM at Néel temperatures of \textit{$T_{N1}$}= 11.5 K and \textit{$T_{N2}$}= 8 K, however they are very close to each other compared to the transitions in DyAlGe, which indicates that there is a a difference in the exchange interactions between these two systems. The Curie-Weiss fits to the inverse magnetic susceptibility in the paramagnetic (PM) region for magnetic fields applied parallel to the three principal crystallographic axes are shown in the respective insets of Fig.~\ref{fgr:MvsT_Ho}. The experimentally determined effective magnetic moment is comparable to the theoretical free-ion value of 10.60~$\mu_{\mathrm{B}}$ for the Ho$^{3+}$ ion. The negative values of $\theta_{\mathrm{P}}$ obtained from the Curie-Weiss fits indicate that antiferromagnetic exchange interactions dominate along all three directions. Notably, the most negative value of $\theta_{\mathrm{P}}$ for $B \parallel a$ suggests anisotropic exchange interactions arising from their directional dependence, similar to that observed in DyAlGe. Furthermore, the magnetic susceptibility exhibits pronounced anisotropy with respect to the applied magnetic field, with the strongest anisotropic response occurring for $B \parallel a$. To further investigate the magnetic anisotropy, we measured the temperature-dependent magnetic susceptibility, $\chi(T)$, under applied magnetic fields up to 7~T along all three crystallographic directions. Similar to DyAlGe, the transition temperature $T_{N1}$ shows the most pronounced field dependence for $B \parallel a$. However, in contrast to DyAlGe, where $T_{N2}$ remains nearly unchanged for $B \parallel c$, HoAlGe exhibits a noticeable shift of $T_{N2}$ with increasing magnetic field along this direction. These observations suggest that the anisotropic exchange interactions and crystal electric field (CEF) effects differ significantly between the two compounds, leading to distinct magnetic phase evolution under applied magnetic fields \cite{MAJUMDAR2001645}.\\
Similar to DyAlGe, we have measured isothermal magnetization at selected temperatures up to 20 K. Interestingly, MMTs with hysteresis between the up and down sweeps of magnetic fields are observed for all field directions,unlike in DyAlGe. Fig.~\ref{fgr:MvsB_Ho}(a) shows magnetization as a function of the field at different temperatures between 1.8-20~K for B $\parallel$ a, where MMTs at 1.8~K with significant hysteresis are observed on the slopes of the $M(B)$ curves near 1.7~T, 2.7~T, and 3.5~T, whereas B $\parallel$ b and B $\parallel$ c show such MMTs above 5~T of magnetic field. The observation of these metamagnetic transitions along directions other than (B $\parallel$ a) in HoAlGe, in contrast to DyAlGe, further highlights the distinct magnetic behavior arising from the different 4f-electron configurations in these systems. This first-order-like magnetic transition gradually gets smeared with increasing temperature and shows reversible $M(B)$ curves above 9 K which is above $T_{N2}$. These different field-induced transitions along different directions suggest the presence of more complex magnetic behavior at higher magnetic fields. Compared to DyAlGe, the $M(B)$ curve measured at 1.8~K for $B \parallel a$ reaches saturation early around 4~T with a similar spontaneous magnetic moment of about $9~\mu_{\mathrm{B}}$/Ho$^{3+}$. In contrast, the magnetization does not reach saturation even at the highest applied field other directions, resulting in substantially lower magnetic moments along these directions. Considering the results for both systems, the pronounced anisotropy in the saturation magnetization indicates strong magnetic anisotropy, with the $a$-axis serving as the magnetic easy axis. While detailed inelastic neutron scattering experiments are required to directly probe the role of anisotropic exchange interactions in the observed metamagnetic transitions, the magnetic structures below $T_{N1}$ and $T_{N2}$ are likely to be different. Furthermore, the crystal structure features an isosceles triangular arrangement of $R$ atoms along the $a$-axis, perpendicular to the $(0kl)$ planes, which may contribute to the unusual metamagnetic behavior observed only for $B \parallel a$. Detailed magnetic structure studies are required to understand the nature of these changes across the lower transition temperature and to confirm the presence of a first-order phase transition, which may be associated with a change in the magnetic propagation vector.


\begin{figure}[t]
	\adjincludegraphics[height=14cm,trim={1.5cm 0.2cm 1.1cm 0cm},clip]{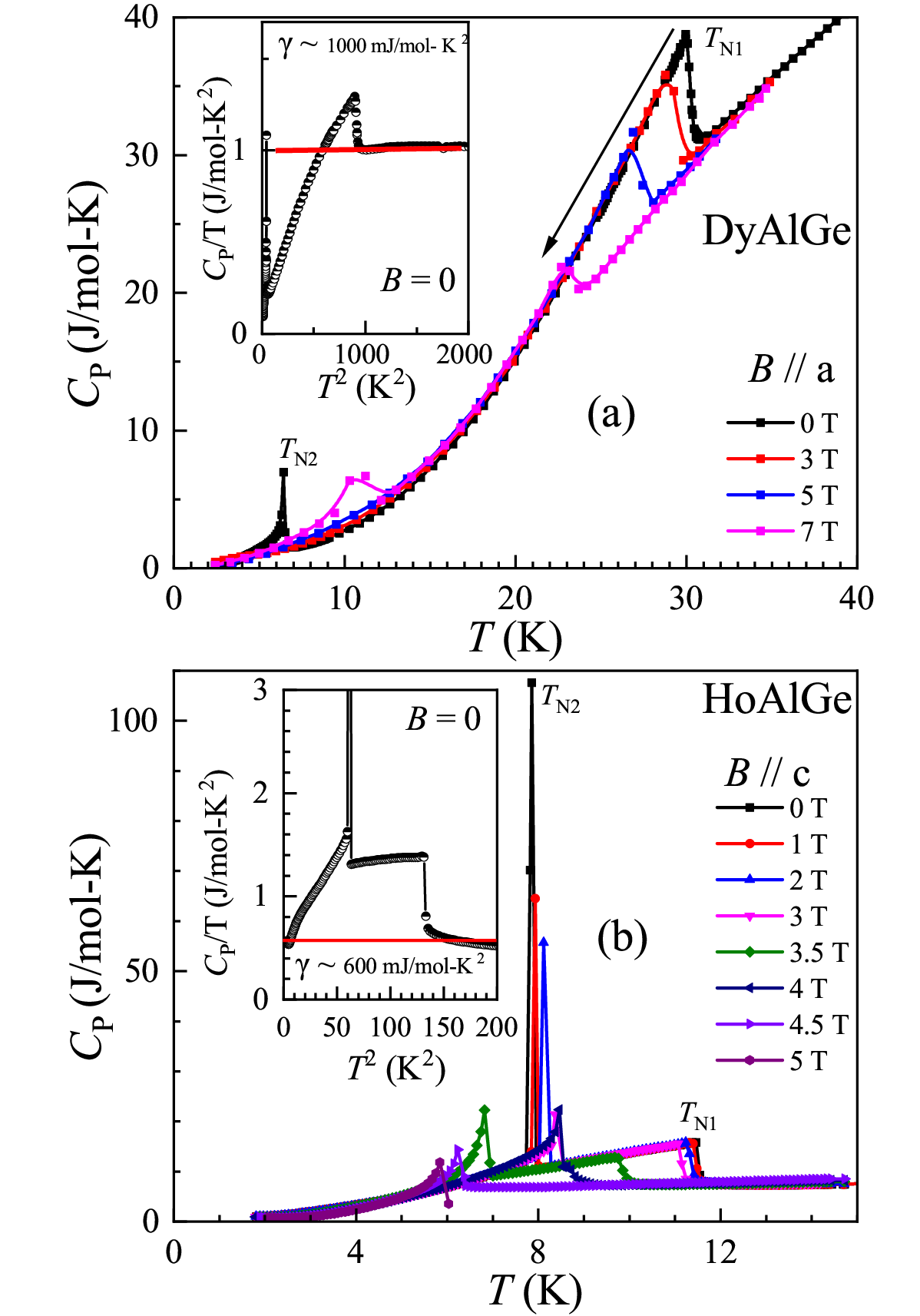}
    \caption{Temperature dependence of Heat Capacity measured under zero field and applied magnetic fields for DyAlGe (a) and HoAlGe (b). The respective insets show the C$_P$/T vs T$^2$ plot in zero field with red solid line showing the liner fit to calculate the $\gamma$. The main panels of C$_P$ vs T plots with different applied magnetic fields shows the variation of $T_{N1}$ and $T_{N2}$ with applied magentic fields.}
    \label{fgr:Cp}
 \end{figure}


\begin{figure}[t]
	\adjincludegraphics[height=12.5cm,trim={0cm 0cm 0cm 0cm},clip]{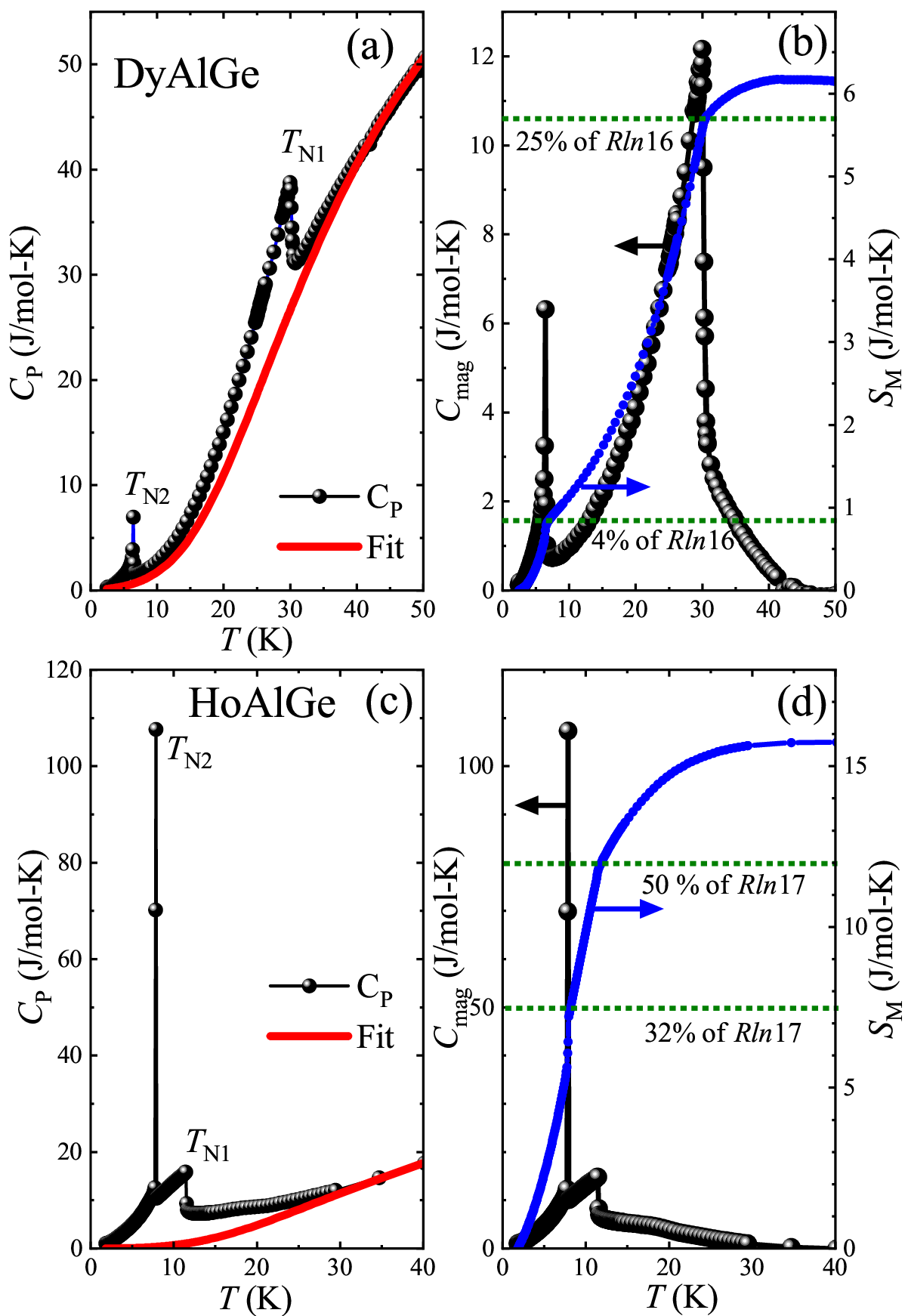}
    \caption{Black curve indicates the experimental temperature-dependent heat capacity, C$_P$(T) and red curve shows the fitting of the experimental curve using the electronic and lattice contributions for both DyAlGe (a) and HoAlGe (c). Temperature-dependent magnetic contribution of heat capacity (C$_{mag}$(T)) and magnetic entropy, S$_M$(T) are presented by black and blue curves in (b) and (d) for DyAlGe and HoAlGe, respectively.}
    \label{fgr:S_combo}
 \end{figure}


 \begin{figure}[t]
	\adjincludegraphics[height=12.2cm,trim={0cm 1cm 0cm 0cm},clip]{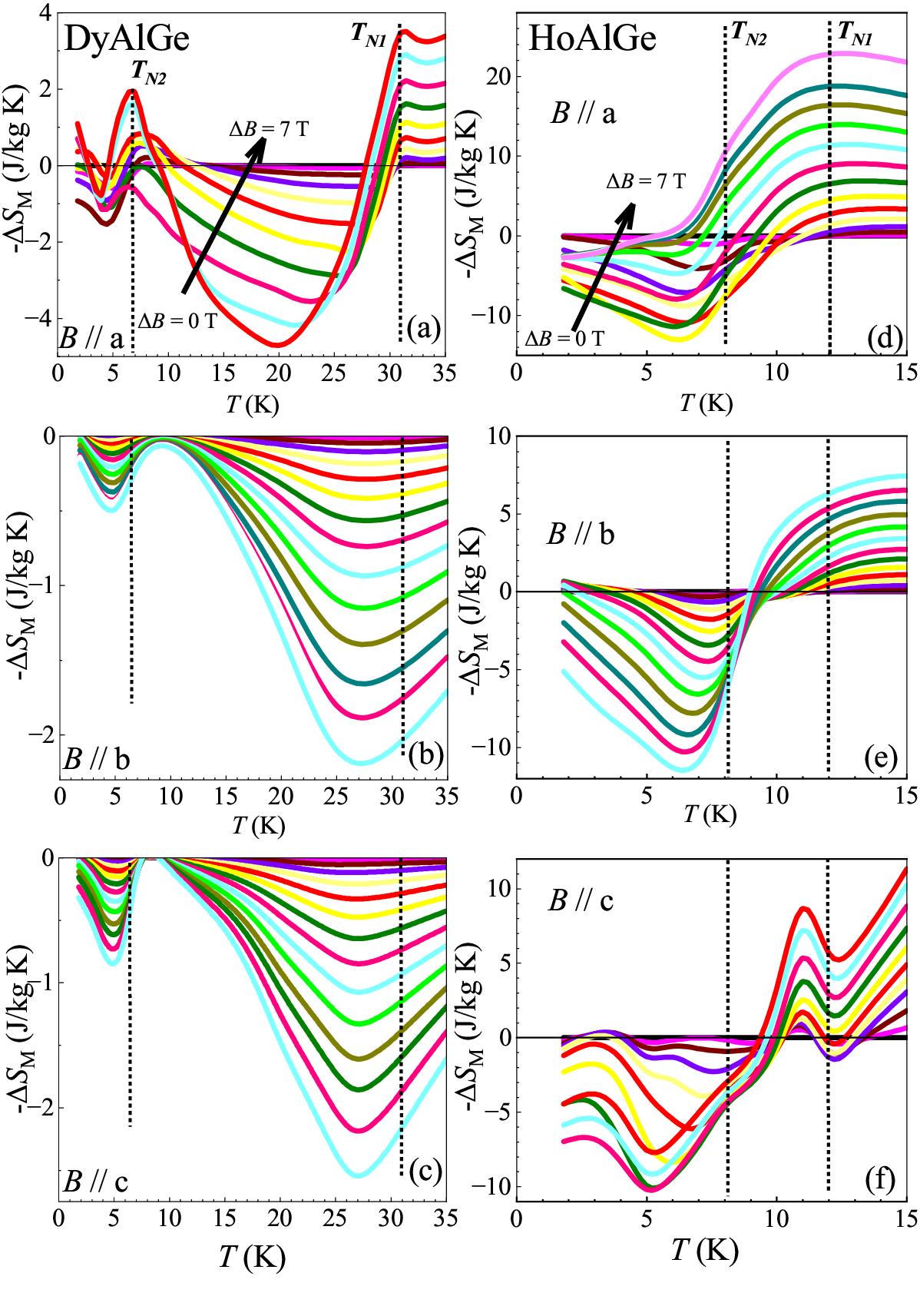}
    \caption{Change in magnetic entropy (-$\Delta$S$_M$) of DyAlGe (a-c) and HoAlGe (d-f) as a function of temperature at different magnetic fields from 0 T to 7~T. Unlike DyAlGe, HoAlGe exhibits a sign change in -$\Delta$S$_M$ for all field directions, suggesting competing magnetic interactions and the evolution of distinct magnetic phases with temperature and magnetic field.}
    \label{fgr:MCE}
 \end{figure}


\subsection{Heat Capacity and Magnetocaloric Effect}

Fig.~\ref{fgr:Cp} shows the zero-field and in-field heat capacity, $C_{P}(T)$, of DyAlGe (a) and HoAlGe (b) as a function of temperature. The data ultimately exhibit two pronounced anomalies at \(T_{N1}\) and \(T_{N2}\), consistent with distinct bulk phase transitions. For DyAlGe, a well-defined $\lambda$-type anomaly is observed at $T_{N1} = 30.5$~K, followed by a second transition at $T_{N2} = 6.6$~K, consistent with the magnetic measurements. Similarly, HoAlGe exhibits an anomaly at $T_{N1}= 11.5$~K, followed by a sharp, $\delta$-function-like transition at $T_{N2}= 8$~K. Notably, the heat-capacity peak at $T_{N2}$ is nearly an order of magnitude larger than that at $T_{N1}$, in stark contrast to the behavior observed in DyAlGe. It should also be noted that a clear thermal hysteresis between the field-cooled cooling (FCC) and field-cooled warming (FCW) magnetic susceptibility curves is observed around $T_{N2}$, providing further evidence for the first-order nature of this magnetic phase transition. In both compounds, $T_{N1}$ gradually shifts to lower temperatures with increasing applied magnetic field. In contrast, $T_{N2}$ exhibits a non-monotonic field dependence in both systems. More interestingly, the heat-capacity peak at $T_{N2}$ in HoAlGe is strongly suppressed even by an applied field of 1~T. The rapid reduction of the sharp $\delta$-like anomaly suggests that the applied magnetic field stabilizes the low-temperature magnetic phase, thereby reducing the entropy change associated with the transition. This field-induced suppression of the heat-capacity peak, together with the thermal hysteresis observed between the field-cooled cooling (FCC) and field-cooled warming (FCW) magnetic susceptibility curves, provides strong evidence that the lower magnetic transition in HoAlGe is first order. The pronounced field sensitivity further suggests a field-induced modification of the magnetic structure, possibly associated with a change in the magnetic propagation vector. The Sommerfeld coefficient, $\gamma$, obtained from linear fits to the low-temperature $C_P/T$ versus $T^2$ data [see the respective insets of Fig.~\ref{fgr:Cp}], is relatively large, suggesting enhanced electronic correlations in these compounds.


 \begin{figure}[t]
	\adjincludegraphics[height=10cm,trim={0.15cm 3cm 0cm 0.1cm},clip]{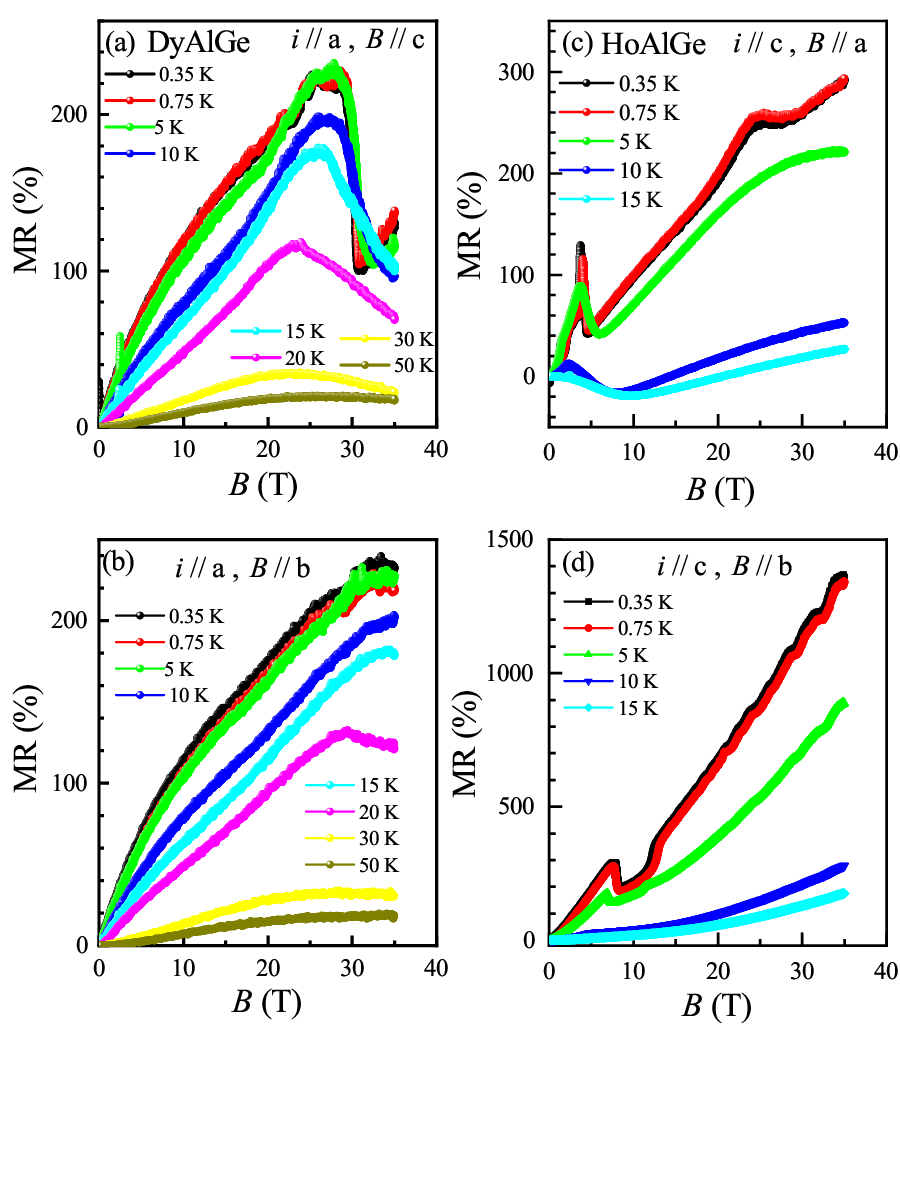}
    \caption{High-field magnetoresistance up to 35 T for DyAlGe [(a), (b)] and HoAlGe [(c), (d)] at different temperatures near the magnetic phase transitions. HoAlGe exhibits more pronounced anisotropic magnetotransport features compared with DyAlGe.}
    \label{fgr:HFMR}
 \end{figure}


The magnetic entropy, S$_M$(T) was obtained by integrating (C$_{mag}$(T)/T) with respect to temperature as shown in Fig. ~\ref{fgr:S_combo}, where C$_{mag}$(T) represents the magnetic contribution to the heat capacity after subtracting the phonon background using the expression given in Ref.\cite{PhysRevB.43.13137, lipika2026strongmetamagneticswitchinganisotropic}. For DyAlGe, the entropy recovered within the measured temperature range is $\approx$ 5.7~J.mol$^{-1}$.K$^{-1}$, corresponding to about 25$\%$ of the theoretical free-ion value, R $\ln$ 16 = 23.05~J.mol$^{-1}$.K$^{-1}$ for Dy$^{3+}$ with \textit{J}=15/2. In HoAlGe, the recovered entropy reaches $\approx$ 11.88 J.mol$^{-1}$,K$^{-1}$, or 50$\%$ of the theoretical value, R $\ln$ 17 = 23.43 J.mol$^{-1}$,K$^{-1}$ for Ho$^{3+}$ with \textit{J}=8. The reduced entropy recovery in both compounds can be understood in terms of crystal-electric field effects, which lift the degeneracy of the (\textit{J}) multiplet and split it into a series of CEF levels. Consequently, the full free-ion entropy is not necessarily released near the magnetic ordering temperatures: the low-lying CEF states may dominate the low-temperature magnetic response, while higher-lying levels contribute additional entropy only when they become thermally populated at elevated temperatures \cite{kjct-68sj,5bbt-q2t7}. The different fractions of recovered entropy in DyAlGe and HoAlGe may therefore reflect differences in their CEF level schemes and the energy scales governing thermal population of excited states. Further spectroscopic or inelastic-neutron-scattering measurements would be needed to establish the CEF level energies and quantify their contributions to the magnetic entropy.\\

We have studied the magnetic entropy change (-$\Delta$S$_M$) in both DyAlGe and HoAlGe, along all three axes to get more information about the different magnetic phase transitions, as studied in other compounds \cite{PhysRevMaterials.5.054407, 10.1063/1.5121293, PhysRevB.101.144440, 10.1063/1.4903770, PhysRevB.80.134422, Fang2017, PhysRevB.65.094405}. We have measured many isothermal M(B) curves for different temperatures with close intervals up to 35~K for DyAlGe and 15~K for HoAlGe, to calculate the magnetocaloric effect (MCE) using the Maxwell equation \cite {tishin2016magnetocaloric}. We have derived the isothermal entropy change, defined as -$\Delta$S$_M$, and the result obtained is shown in Fig.~\ref{fgr:MCE} for both of the systems. Notably, for $B \parallel a$, the sign of -$\Delta$S$_M$ is flipped twice below the magnetic ordering, (i) around 30~K, which indicates a strong competition among the magnetic interactions and implies the dominance of an antiferromagnetic component over the ferromagnetic (FM) phase \cite {PECHARSKY199944}, and (ii) at low temperature just below \textit{$T_{N2}$}, where possibly a first-order magnetic phase transition exits; at higher temperatures, the sign of -$\Delta$S$_M$ is positive, which is a signature of a tendency for spin-reorientation \cite {LI2020153810}, suggesting the dominance of ferromagnetic correlations in the presence of such magnitudes of external fields. Interestingly, other directions depict the peaks at magnetic transition temperatures, also with lower value of -$\Delta$S$_M$ compared to $B \parallel a$. In contrast to DyAlGe, HoAlGe has a more anisotropic MCE for magnetic field applied along all three axes. However, the highest value of -$\Delta$S$_M$ is achieved only for $B \parallel a$ similar to DyAlGe. Compared to DyAlGe, HoAlGe has a stronger signature for the presence of a first-order magnetic transition near 8~K, manifested through magnetic hysteresis and corroborated by a very high sharp peak in heat capacity data Figs.~\ref{fgr:Cp}. Notably, the isothermal entropy change around \textit{$T_{N1}$} has the largest peak value for $B \parallel a$ ; for a field change from zero to 7~T, the entropy change is ~3.5 J/kg K (DyAlGe) and ~23 J/kg K (HoAlGe) suggesting a role of anisotropy of 4f orbital in determining this large value for HoAlGe \cite{magnetochemistry9030085, 10.1063/5.0016724, IYER2023170406}. The multiple sign reversals together with overall high values of magnetic entropy point to a significant magnetocaloric potential in these magnetic systems.

   \begin{figure}[t]
	\adjincludegraphics[height=15cm,trim={0.2cm 0.2cm 0cm 0cm},clip]{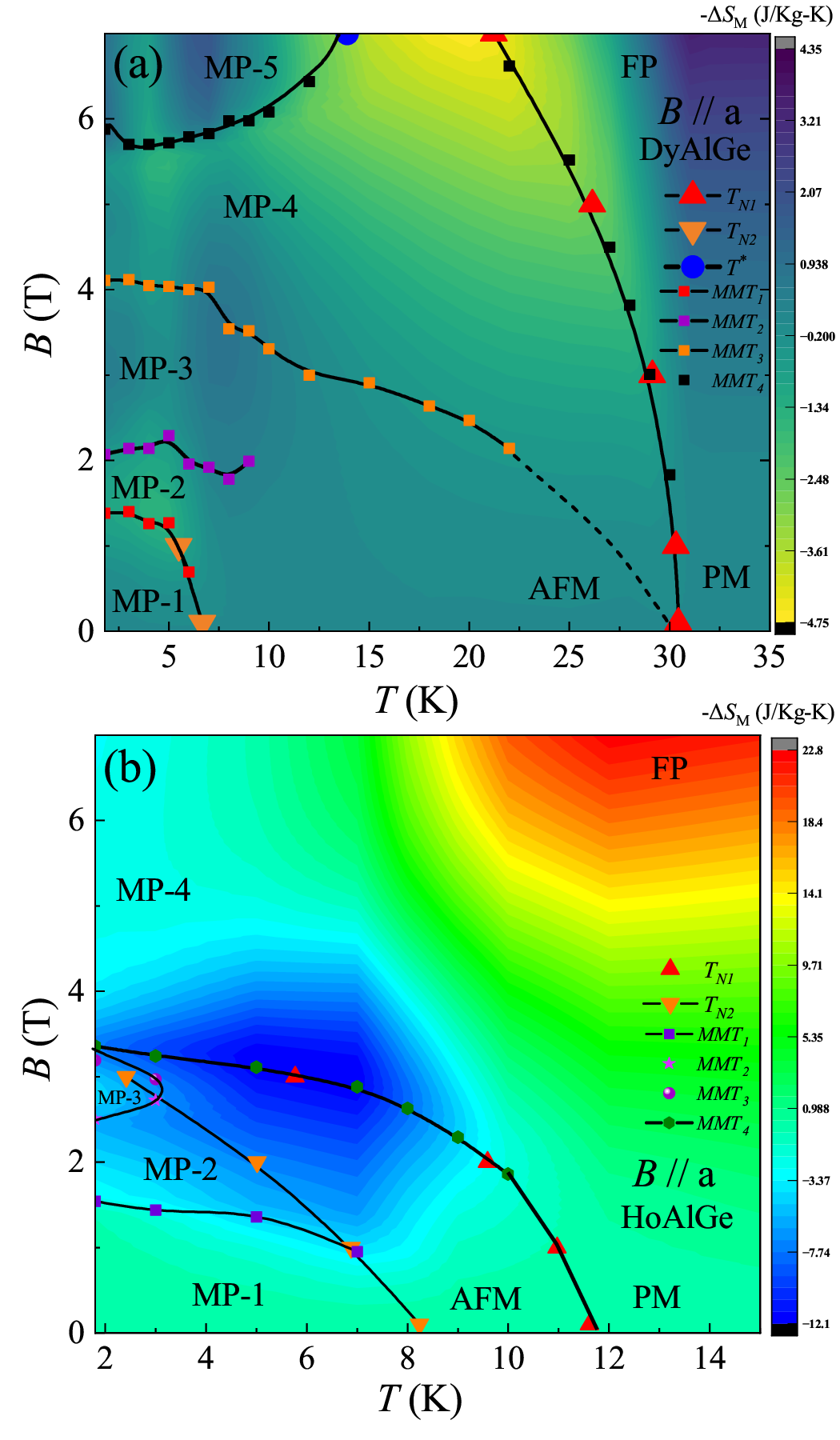}
    \caption{\textit{B}-\textit{T} phase diagram of both DyAlGe (a) and HoAlGe (b) as a contour plot of magnetic entropy change, with  \textit{$T_{N1}$} and \textit{$T_{N2}$} derived from magnetic susceptibility measurements. The different metamagnetic transitions are derived from the respective magnetization as a function of field.}
    \label{fgr:Phase Diagram}
 \end{figure}


\subsection{Magnetotransport Properties up to high magnetic field of 35 T}
We have measured electrical transport as a function of temperature and field along interesting directions for both DyAlGe and HoAlGe. The residual resistivity ratio (RRR), calculated as ${\rho}$(300 K)/${\rho}$(1.8 K), shown in Fig.~\ref{fgr:RRR}, indicates fair crystal quality for both of these grown crystals. Although the general trend of electrical resistivity ${\rho}$ as a function of temperature in the paramagnetic state of both these systems is metallic, an anomalous behavior is observed on approaching the respective \textit{T$_{N1}$} from higher temperature (1.8-50 K) which is consistent with the recently reported transport behavior in these systems \cite{SLADE2026174537}. Similar to the magnetic measurements, the respective\textit{T$_{N1}$} and \textit{T$_{N2}$} varies more prominently for \textit{$B \parallel a$} compared to the other directions. To further explore the existing magnetic phases at higher fields, we have measured the magnetoresistance (MR). The observation of multiple metamagnetic transitions for (\textit{$B \parallel a$}), consistent with the magnetization data, highlights the pronounced magnetic anisotropy of these compounds. The $a$-axis thus emerges as the most favorable direction for probing field-induced changes in electronic transport, as successive magnetic transitions can modify the coupling between localized 4\textit{f} moments and conduction electrons, potentially giving rise to distinct magnetotransport signatures. To further investigate these field-induced phenomena, we performed high-field magnetotransport measurements, as shown in Fig. 10. For DyAlGe, while no metamagnetic transitions are observed for (\textit{$B \parallel b$}) and (\textit{$B \parallel c$}) up to 14 T, additional transitions emerge along both directions at fields above 20 T. HoAlGe exhibits even more intriguing high-field behavior. For (\textit{$B \parallel a$}), in addition to the low-field MMTs consistent with those observed in the magnetization measurements, an additional transition appears above 20 T. Furthermore, for (\textit{$B \parallel c$}), clear quantum oscillations emerge at fields above 18 T. A preliminary analysis suggests three dominant oscillation frequencies at approximately 131, 314, and 576 T. Although these observations provide an initial indication of the underlying Fermi-surface topology, additional high-field measurements and a more detailed analysis are needed to establish the precise Fermi-surface characteristics. These high-field measurements reveal additional field-induced features beyond those accessible in conventional magnetization measurements, providing further insight into the complex magnetic phase diagrams and the interplay between magnetism and electronic transport in DyAlGe and HoAlGe.\\

\subsection{Magnetic Phase Diagram}

Detailed results of both these single crystals of DyAlGe and HoAlGe can be summarized in the magnetic phase diagrams as shown in Fig.~\ref{fgr:Phase Diagram}, measured down to 1.8~K and up to an applied magnetic field of 7~T. Contour plots of magnetic entropy, with the different data points marking the transition temperatures extracted from magnetic susceptibility, and also from \textit{B}(H) curves at different temperatures are shown in Fig.~\ref{fgr:Phase Diagram} for both DyAlGe (a) and HoAlGe (b). The dotted lines are guides to eye to show possible boundaries from color contrast in the contour plots. It can be seen that the dominant AFM phase just below the respective magnetic ordering temperatures evolves into multiple magnetic phases under high applied fields at low temperatures. At sufficiently high fields, both systems exhibit a field-polarized (FP) phase. DyAlGe and HoAlGe exhibit multiple low-field magnetic phases at low temperatures, similar to those observed in TbAlGe. However, the temperature ranges and magnetic transition temperatures associated with these phases differ among the three compounds, reflecting the distinct magnetic interactions and anisotropies arising from the different rare-earth ions. These phase diagrams highlight the fragility of the magnetic phases below the second magnetic transition, underscoring the nontrivial interplay of competing magnetic interactions in these systems. Although the various low-temperature phases exhibit a sequence of magnetic transitions, the penultimate transition below the respective second magnetic transition appears to be strongly first-order, as evidenced by the hysteresis observed in the M(B) curves, whereas the remaining transitions appear to be continuous. This behavior is also reflected in the magnetic entropy plots, where the transitions in the low-temperature and low-field region are clearly manifested by sign changes across the distinct magnetic phases. Sharing the same crystal structure as TbAlGe, these systems exhibit similar magnetism, likely arising from anisotropic exchange interactions and magnetic frustration \cite{PhysRevB.78.064412, PhysRevB.100.094442, PhysRevB.101.144440, 10.1063/1.5121293, 10.1103/physrevb.60.12162, kurumaji2019skyrmion, 1sc3-6cfp, annurev:/content/journals/10.1146/annurev.ms.24.080194.002321}. Such interactions may stabilize nontrivial spin structures and highlight these centrosymmetric orthorhombic systems as potential platforms for Weyl fermions, with magnetism strongly influencing their topological electronic structure.

\section{Conclusion}
In summary, we have grown high-quality single crystals of heavy rare-earths DyAlGe and HoAlGe, and performed detailed structural, magnetic and magnetocaloric studies on these magnetic systems, which crystallize in the orthorhombic $Cmcm$ structure. DyAlGe exhibits two antiferromagnetic transitions at $T_{N1}$ = 30.5~K and $T_{N2}$ = 6.6~K, whereas HoAlGe also has two antiferromagnetic transitions at $T_{N1}$ = 11.5~K and $T_{N2}$ = 8~K. Both of these systems develop a rich, field-tuned magnetic phase diagram with multiple metamagnetic transitions for magnetic field parallel to the $a$-axis. Notably, unlike DyAlGe, HoAlGe strongly suggests a first-order transition below the second transition temperature as seen by observed hysteresis both in temperature and field dependent magnetic properties. Similar to the TbAlGe, high-field measurements up to 41.5~T would be interesting to reveal additional metamagnetic transitions beyond those observed in conventional measurements up to 14~T in these systems. The magnetic entropy change  demonstrates a significant MCE in these systems, with a maximum -$\Delta$$S_M$ of $\approx$ 5-23 J/kg-K for a field change of 0-7 T. The sign reversal of -$\Delta$$S_M$ across field-induced transitions highlights the complex field-tuned spin configurations. The moderate refrigeration capacity and magnetic anisotropy suggest that these materials may serve as model systems for studying magnetocaloric effects in topologically nontrivial intermetallics. Remarkably, the detailed phase diagrams for magnetic field parallel to $a$-axis for both DyAlGe and HoAlGe suggest that the multiple magnetic phases exist in a subset of these metamagnetic transitions. Altogether, these results provide key insight into the interplay between magnetism in both of these systems, making it an ideal platform to study correlated physics. Future studies combining band-structure calculations and neutron scattering experiments will be essential to fully elucidate the existing magnetic phases in these systems.

\begin{acknowledgments}

Research at the University of Maryland was supported by the Gordon and Betty Moore Foundation’s EPiQS Initiative through Grant No. GBMF9071, the U.S. National Science Foundation (NSF) Grant No. DMR2303090, the Binational Science Foundation Grant No. 2022126, and the Maryland Quantum Materials Center. R.K. acknowledges support from the National Institute of Standards and Technology Cooperative Agreement 70NANB17H301. A portion of this work was performed at the National High Magnetic Field Laboratory, which is supported by National Science Foundation Cooperative Agreement No. DMR-2128556 and the State of Florida. 
\end{acknowledgments}

\nocite{*}

\bibliography{RAlGe}

\end{document}